\documentclass[a4paper,12pt,notitlepage]{article}
\usepackage{graphicx} \usepackage{amsmath} \usepackage{amssymb}
\usepackage{psfrag}  \usepackage{color,xcolor}
\usepackage[latin1]{inputenc} 

\usepackage{natbib}
\usepackage{braket}

\usepackage{geometry}
\usepackage[colorlinks=true,linkcolor=blue,citecolor=blue]{hyperref}

\usepackage{authblk}
\newcommand{\be}{\begin{equation}}
\newcommand{\ee}{\end{equation}}
\newcommand{\bs}{\boldsymbol}

\begin{document} 

\title{Photonic quantum Hall effects}
\author[1]{Daniel Leykam}
\author[2,\thanks{Corresponding author: daria.smirnova@anu.edu.au}]{Daria Smirnova}
\affil[1]{Centre for Quantum Technologies, National University of Singapore, 3 Science Drive 2, Singapore 117543}
\affil[2]{Nonlinear Physics Centre, Research School of Physics, Australian National University, Canberra ACT 2601, Australia} 
\date{}
% \date{\today}
\maketitle
\begin{abstract}
This article reviews the development of photonic analogues of quantum Hall effects, which have given rise to broad interest in topological phenomena in photonic systems over the past decade. We cover early investigations of geometric phases, analogies between electronic systems and the spectra of periodic photonic media including photonic crystals, efforts to generalize topological band theory to open, dissipative, and nonlinear wave systems, pursuit of useful device applications, and ongoing studies of photonic Hall effects in  classical nonlinear optics and the quantum regime of strong photon-photon interactions.
\end{abstract}

\textbf{Keywords: geometric phase; spin Hall effect; band structure; topological phase; edge states; non-Hermitian; nonlinear optics; metamaterial; photonic crystal} 

\subsection*{Key Points / Objectives Box}

\begin{itemize}
    \item Review theoretical and experimental work on topological phases of light
    \item Emphasize similarities and differences compared to electronic quantum Hall effects
    \item Discuss possible device applications of photonic quantum Hall effects, including robust waveguides and lasers 
    \item Outline new developments in the study of classical and quantum photonic topological systems
\end{itemize}

\section{Introduction}

The quantum Hall (QH) phase discovered in 1980, which arises in two-dimensional electron gases under a strong applied external out-of-plane magnetic field, represents the first example of a topological insulating phase of matter~\citep{Klitzing2020}. Such materials are distinguished by bulk band gaps in their band structures with gapless chiral edge states which propagate unidirectionally along the boundaries with a built-in topological immunity to backscattering from disorder. 

The use of topological concepts to explain robust physical observables originated with the seminal works of Thouless \textit{et~al.}~\citep{Thouless1982, Kohmoto1985} connecting the 
experimentally observed quantized Hall conductance with a topological invariant, namely, the Chern number, defined in the momentum space. Studies of related topological insulating materials have attracted enormous interest from the mid 2000s until the present day. 

Interestingly, topological phases are not limited to fermionic systems and can be translated to classical wave phenomena. Unusual manifestations inherent to topologically nontrivial states, including the ability of edge modes to overcome structural imperfections without back-reflection, drive general interest in topological and quantum Hall effects of classical waves, especially within photonics and optical communications~\citep{Ozawa2019,Smirnova2020APR}. 

Topological photonics has recently emerged as a novel approach to robust waveguiding and routing of light. It exploits engineered photonic structures~\citep{Lu2016} with properties analogous to electronic topological insulators~\citep{Hasan2010}, which are insulating in their bulk but exhibit conducting states at their surfaces.

While discussions of photonic topological phases are usually framed as being inspired by and analogous to more widely-known electronic topological phases, closer inspection of the history of electronic and photonic topological effects reveals they are in fact closely intertwined. For example, the explanation of the quantum Hall effect in terms of the Chern number was quickly found to be related to geometric phases anticipated decades earlier in optics. 

Here we will attempt to present a complete history of the study of photonic quantum Hall effects from its surprisingly early origins up to ongoing research. To keep this Chapter concise we will focus on two-dimensional photonic quantum Hall and spin-Hall effects, mentioning related topological phenomena (e.g. one-dimensional edge states, higher-order topological phases) only in passing and directing the interested reader to relevant in-depth reviews.

\section{Overview}

\subsection{Historical Background}

The study of photonic quantum Hall effects can be loosely divided into three eras: 
\begin{enumerate} 
\item A pre-quantum Hall era of foundational theoretical and experimental works on optical beam and polarization shifts, now understood in terms of geometric phases.
\item A period of growing interest in analogies between condensed matter physics and photonics,
culminating in the first observation of topological photonic edge states at microwave frequencies.
\item The topological photonics era, sparked by the crucial breakthrough that symmetry protection could circumvent the need for magneto-optic materials, opening up the exploration of topological phenomena to a wider variety of photonic systems.
\end{enumerate}

\subsubsection{Pre-quantum Hall era (1938--1980)}

\begin{figure}
    \centering
    \includegraphics[width=\columnwidth]{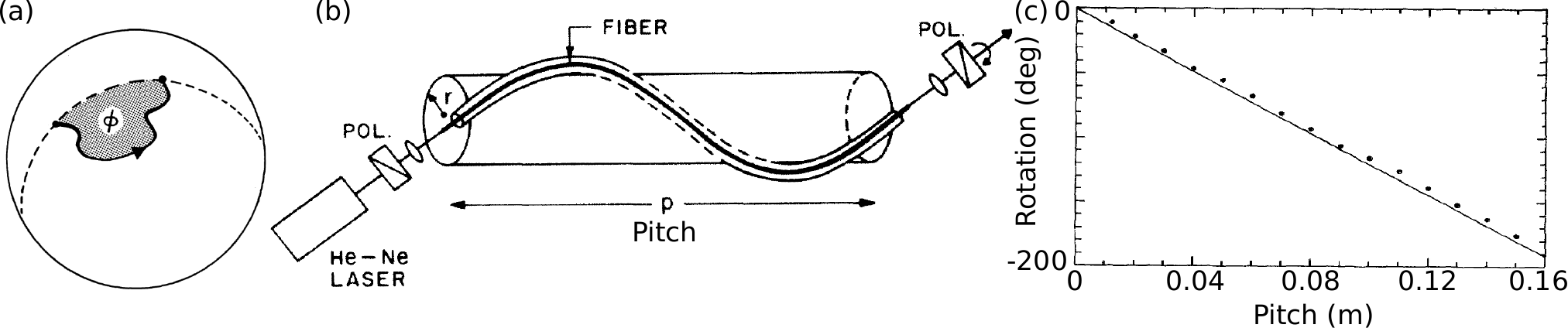}
    \caption{Geometric phase in twisted optical fibers. (a) The geometric polarization rotation is determined by the solid angle $\phi$ subtended the fiber trajectory. Adapted from \cite{Haldane:86}. (b) Experimental setup for measuring the geometric polarization rotation, adapted from \cite{PhysRevLett.57.937}. (c) Polarization rotation induced by a single twist as a function of its pitch, adapted from \cite{Ross1984}. Points are experimental measurements and the solid line is the prediction from theory. }
    \label{fig:geometric_phase}
\end{figure}

The Chern number underling the quantum Hall effect is closely related to the parallel transport properties of quantum states, which form rays in Hilbert space. Already in 1938 Rytov found that classical rays of light undergo a polarization rotation under non-planar propagation~\citep{rytov1938transition}, shown in Fig.~\ref{fig:geometric_phase} for the case of light propagation in a twisted optical fiber. The polarization rotation is determined purely by the area enclosed by the trajectory, being an example of a geometric phase~\citep{vladimirskii1941rotation,Pancharatnam1956}. Another foundational discovery was the Imbert-Federov shifts of reflected light beams~\citep{fedorov1955k,PhysRevD.5.787}, which are now recognized as an example of spin Hall effects of light~\citep{Bliokh_2013}. 

\subsubsection{Quantum Hall era (1980 -- 2011)}

Shortly after Thouless and collaborators explained the quantum Hall effect in terms of a topological invariant~\citep{Thouless1982}, \cite{PhysRevLett.51.2167} showed how the invariant was related to the quantum geometric phase studied by Berry~\citep{doi:10.1098/rspa.1984.0023}. While the first experimental observation of Berry's geometric phase as polarization rotation in a twisted optical fiber loop \citep{PhysRevLett.57.937} argued it was a genuinely quantum mechanical effect that survives in the classical limit \citep{PhysRevLett.57.933}, \cite{Haldane:86} emphasized that this and earlier polarization rotation experiments by \cite{Ross1984} could be understood purely classically using differential geometry. Nevertheless, these geometric phase experiments and theoretical proposals for photonic crystals inspired by electronic band gaps \citep{PhysRevLett.58.2059,PhysRevLett.58.2486} demonstrate the increasing cross-fertilization of ideas between quantum condensed matter physics and optics. 

\cite{PhysRevLett.61.2015} presented a condensed matter lattice model exhibiting a quantum Hall effect in the absence of a magnetic field, demonstrating how time reversal symmetry-breaking, not magnetic flux, underlies the quantum Hall effect. While the broad importance of this finding was not widely appreciated at the time (the article attracted only a few dozen citations over the next decade), Haldane's model formed the basis for the surge in interest in topological materials in the 2000s~\citep{Hasan2010}. 

The 1990s saw continued interest in geometric phase effects, including their influence on the semiclasical dynamics of wavepackets in periodic potentials studied by Niu -- a former PhD student of Thouless -- and his collaborators~\citep{PhysRevLett.75.1348,RevModPhys.82.1959}.
%~\citep{PhysRevLett.75.1348,PhysRevB.53.7010,PhysRevB.59.14915}
%, reviewed by~\cite{RevModPhys.82.1959}. 
The semiclassical formalism showed how weak intrinsic Hall and spin Hall effects could be strongly enhanced in certain periodic potentials, inspiring further studies of spin Hall effects of electrons~\citep{doi:10.1126/science.1087128,doi:10.1126/science.1105514} and photons~\citep{PhysRevLett.93.083901,PhysRevE.74.066610, BLIOKH2004181,PhysRevE.70.026605,PhysRevLett.95.136601,Leyder2007}.

\cite{PhysRevLett.95.226801} showed that strong spin-orbit coupling may imitate the effect of a magnetic field in time-reversal invariant electronic systems, giving rise to a quantum spin Hall effect. Its simplest form constitutes essentially two time-reversed copies of the quantum Hall phase, where up and down spin electrons are decoupled and experience opposite effective magnetic fields. Remarkably, the quantum spin Hall effect is robust to coupling between the two spin sectors, forming a distinct topological phase of matter protected by time-reversal symmetry~\citep{QSHE_2005}.

\begin{figure}
    \centering
    \includegraphics[width=0.8\columnwidth]{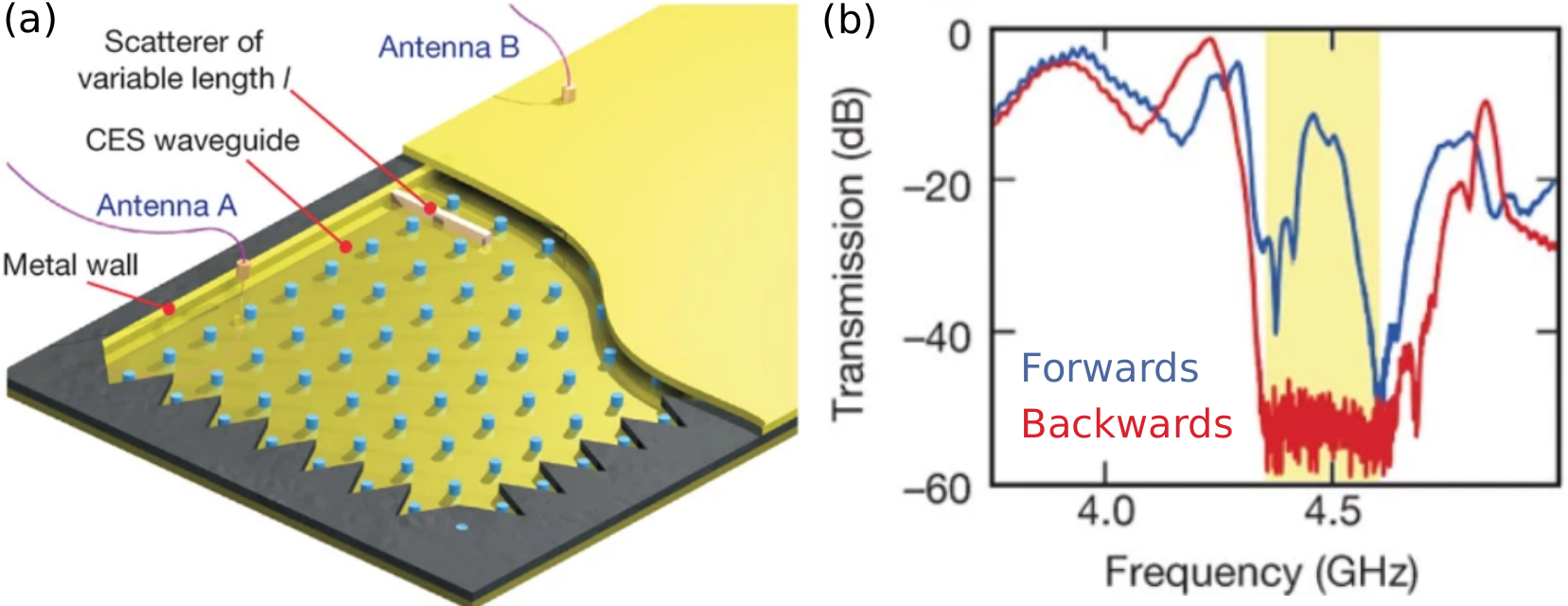}
    \caption{(a) Schematic of two-dimensional magneto-optic photonic crystal including a pair of antennas to measure the transmission of the photonic quantum Hall edge state past a strong scatterer. (b) Measured transmission spectrum from A to B (forward; blue) and from B to A (backward; red). Yellow shaded region denotes the topological band gap exhibiting strong unidirectional transmission. Adapted from \cite{Wang2009}.}
    \label{fig:Wang2009}
\end{figure}

Also in 2005, Haldane and Raghu proposed a photonic analogue of the quantum Hall effect in time reversal symmetry-breaking photonic cyrstals in a preprint that was only published in Physical Review Letters three years later~\citep{Haldane2008}. While potential applications as a basis for broadband unidirectional optical waveguiding were compelling, the weak strength of magnetic effects at optical frequencies suggested this proposal would be challenging to implement in practice. ~\cite{Wang2008} predicted that, even if an optical frequency realization might be impractical, the effect could be readily observed at microwave frequencies. Their seminal experiments using a magneto-optical photonic crystal shown in Fig.~\ref{fig:Wang2009} demonstrated for the first time that robust quantum Hall-like edge states could be observed in photonic systems \citep{Wang2009}.

\subsubsection{Topological photonics era (2011--present)}

Present broad interest in topological bands and quantum Hall effects for photons was sparked by the discovery of several alternatives to weak magneto-optical effects:
\begin{enumerate}
    \item Spin-selective excitations that effectively break the time-reversal symmetry of all-dielectric structures \citep{Hafezi2011}.
    \item Suitably-designed temporal modulation of the coupling in optical resonator lattices that creates effective magnetic fields for light \citep{Fang2012}.
    \item Waveguiding geometries with longitudinal refractive index modulations, where the propagation axis plays the role of ``time'' \citep{Rechtsman2013}.
    \item Metamaterials that emulate the electronic spin-orbit coupling underlying quantum spin Hall phases \citep{Khanikaev2013}.
\end{enumerate}
While these proposals each had their own limitations, (e.g. being limited to weakly-coupled cavity or waveguide lattices), more general and accessible designs soon followed, including all-dielectric topological photonic crystals \citep{Wu2015,Ma2016,Barik2016}. There now exists a wide variety of approaches for generating quantum Hall-like edge stats in photonic systems. 

\subsection{Theoretical and Experimental Approaches}

\label{sec:approaches}

To put photonic quantum Hall effects on a solid theoretical footing it is necessary to demonstrate a correspondence with the models of topological phases studied in condensed matter physics. On the other hand, the differences between Maxwell's equations for electromagnetism and the Schr\"odinger equation provide the opportunity to explore novel classes of topological phenomena inaccessible in electronic condensed matter systems. 

\subsubsection{Schr\"odinger formalism for electromagnetism}

Consider the source-free Maxwell's equations for time $t$ and space $\bs{r}$-dependent electromagnetic fields $\bs{E}(\bs{r},t),\bs{D}(\bs{r},t),\bs{B}(\bs{r},t),\bs{H}(\bs{r},t)$, in units with speed of light $c=1$,
\begin{align} 
   \nabla \times \bs{E} = -\partial_t \bs{B}, \;  \; \nabla \times \bs{H} = \partial_t \bs{D}, \; & \; \nabla \cdot \bs{D} = 0, \; \; \nabla \cdot \bs{B} = 0, \label{eq:maxwell} \\
   \bs{D} = \varepsilon_0 \bs{E} + \bs{P}, \; & \; \mu_0 \bs{H} = \bs{B} - \mu_0\bs{M},
\end{align}
where $\varepsilon_0$ and $\mu_0$ are the permittivity and permeability of free space, respectively, and $\bs{P}(\bs{r},t)$ and $\bs{M}(\bs{r},t)$ are the local polarization and magnetization densities of the material. A Schr\"odinger-like wave equation can be obtained by introducing the ``wavefunction'' $(\bs{E}, \bs{H})^T$ and assuming a linear constitutive relation between the fields, $(\bs{D}, \bs{B})^T = \hat{W}(\bs{E}, \bs{H})^T$, corresponding to $\bs{P}$ and $\bs{M}$ being proportional to the local $\bs{E}$ and $\bs{B}$ fields,
\be 
 i \partial_t \hat{W} \left(\begin{array}{c} \bs{E} \\ \bs{H}\end{array}\right) = \left(\begin{array}{cc} 0 & i\nabla \times  \\ -i\nabla \times & 0 \end{array} \right) \left(\begin{array}{c} \bs{E} \\ \bs{H}\end{array}\right).
\ee
Assuming monochromatic fields with frequency $\omega$, i.e. $(\bs{E}(\bs{r}, t), \bs{H}(\bs{r},t))^T = e^{-i \omega t} \Psi$, and letting $\mathcal{\hat{H}} = (-\nabla \times) \otimes \hat{\sigma}_y$, we obtain the generalized electromagnetic eigenvalue problem
\be 
 \omega \hat{W} \Psi = \mathcal{\hat{H}} \Psi. \label{eq:gep}
\ee
If the constitutive matrix $\hat{W}$ is positive definite the generalized eigenvalue problem is Hermitian and supports a complete set of modes $\Psi_n$ orthonormal with respect to the modified inner product,
\be 
\braket{\Psi_m | \Psi_n} = \int \Psi_m^* \hat{W} \Psi_n d\bs{r} = \delta_{m,n},
\ee 
where $\delta_{n,m}$ is the Kronecker delta function. Therefore periodic electromagnetic media can be studied using standard tools from electronic condensed matter, including the decomposition of the spectrum into a band structure using Bloch's theorem.

There do however exist some peculiar differences between Schr\"odinger and electromagnetic waves. For example, the physical $\bs{E}$ and $\bs{H}$ fields are real-valued, demanding a particle-hole symmetry between positive and negative frequency modes of Eq.~\eqref{eq:gep}; this means the spectrum of $\mathcal{\hat{H}}$ unbounded for both positive and negative frequencies, in contrast to the spectrum of the usual Schr\"odinger equation, which has a ground state. Moreover, the response of a material to an electromagnetic field is in general non-instantaneous; causality (Kramers-Kronig relations) requires the components of the constitutive matrix to be frequency-dependent and complex, i.e. $\hat{W} = \hat{W}(\bs{r},\omega) = \hat{W}^*(\bs{r},-\omega)$. For further more rigorous discussion on the Schr\"odinger formalism for electromagnetism and implications for topological band theory, we recommend the article by \cite{DENITTIS2018579}.

\subsubsection{Routes to topological bands}

\begin{figure}
    \centering
    \includegraphics[width=\columnwidth]{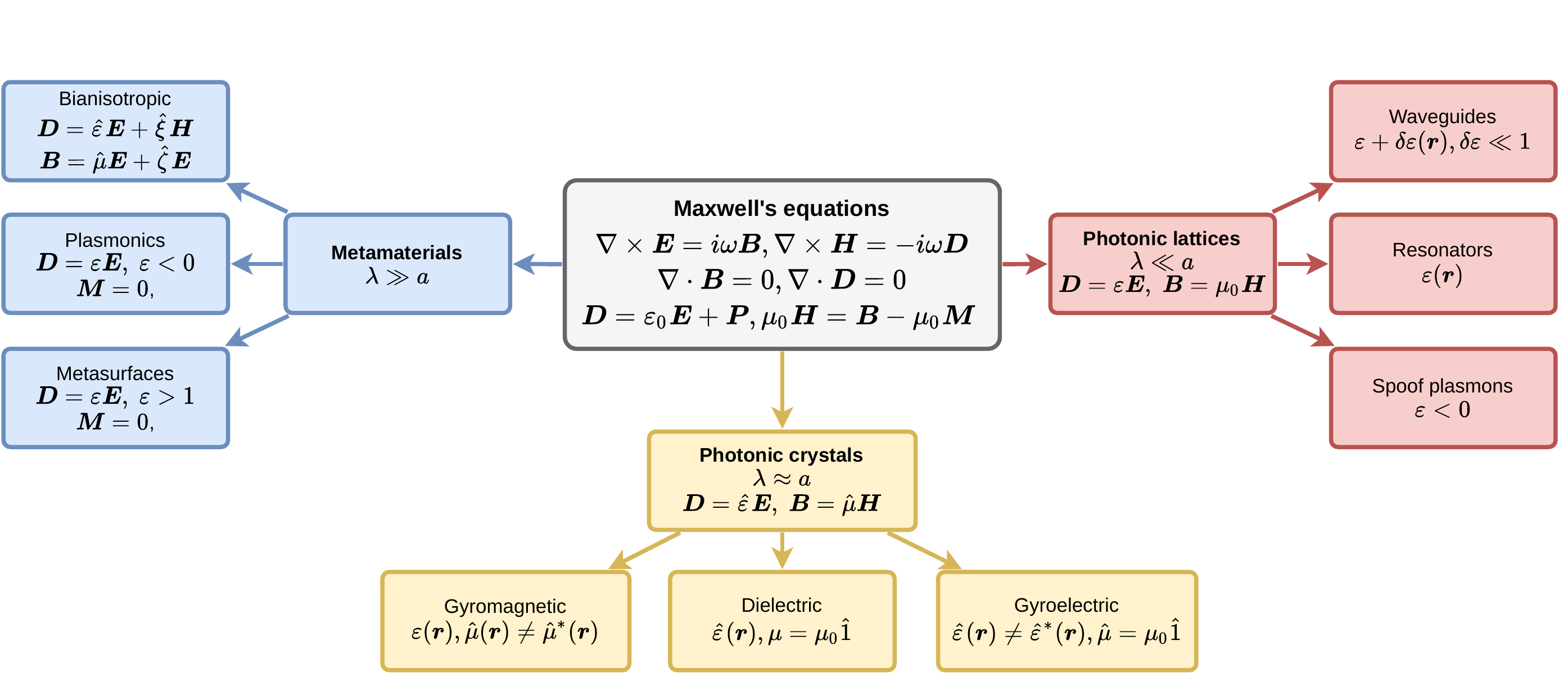}
    \caption{Summary of different approaches for realizing photonic quantum Hall and spin Hall effects, grouped according to the ratio of the operating wavelength $\lambda$ to the modulation scale $a$. Note that generally $a$ is less than structure periodicity $d$.}
    \label{fig:approaches}
\end{figure}

A spatially-periodic constitutive relation plays the role of a periodic potential in the electronic Schr\"odinger equation; appropriate choices of $\hat{W}$ can therefore be used to create topological band gaps supporting photonic quantum Hall states. Different approaches are possible depending on the wavelength of interest $\lambda$ and the scale of structural modifications $a$, summarized in Fig.~\ref{fig:approaches}.

The most ubiquitous topological phases, in terms of characteristic topological invariants, are the Chern insulators characterized by the first Chern number.  For a particular energy band, the Chern number is defined as an integral of the Berry curvature over the first Brillouin zone, $ C_n = \frac{1}{2\pi} \int_{B\!Z} \mathrm{d}^2 \mathbf{k} \, \mathbf{\Omega}_n (\mathbf{k})$.
A non-zero Chern number implies discontinuity of Berry connection and emergence of the topological charge, which leads to the non-vanishing geometric phase term in a eigenvector. For finite systems, the bulk Chern number accounts for the number of unidirectional topologically protected edge modes~\cite{Hasan2010}.
% [Fig. \ref{fig1}(d)]. 

In two-dimensional Chern insulators  
the TR symmetry is necessarily broken, for example, by the magnetic field bias, and in finite Chern TIs the propagation direction of topological edge states is controlled by the sign of the total magnetic flux. 

The original experiments of \cite{Wang2009} broke time-reversal symmetry by applying a magnetic field to a square-lattice photonic crystal 
of gyromagnetic ferrite rods confined vertically between two metallic plates to mimic the TM-polarized modes in infinitely long cylinders, shown in Fig.~\ref{fig:Wang2009}(a). The resultant band structure hosts a gapless chiral edge state that propagates around defects with back scattering % sufficiently 
significantly suppressed and, by contrast, high transmittance within the % second 
band gap shaded yellow in Fig.~\ref{fig:Wang2009}(b). Theoretically, the system is described by periodically-modulated scalar permittivity $\varepsilon(\bs{r})$ and inverse permeability tensor
\be 
\hat{\mu}^{-1}(\bs{r}) = \left(\begin{array}{ccc} \tilde{\mu}^{-1} & i \eta & 0 \\ -i \eta & \tilde{\mu}^{-1} & 0 \\ 0 & 0 & \mu_0^{-1} \end{array} \right),
\ee
where the time reversal symmetry-breaking off-diagonal terms $\pm i \eta$ are induced by an external out-of-plane magnetic field. Eliminating the magnetic field $\bs{H}$ from Maxwell's equations Eq.~\eqref{eq:maxwell} and introducing the ``wavefunction'' $\psi = E_z \sqrt{\tilde{\mu}}$, the eigenvalue problem reduces to \citep{Wang2008}
\be 
-|\nabla + i \bs{A}(\bs{r})|^2 \psi + V(\bs{r}) \psi = 0,
\ee 
where 
\be 
\bs{A} = \frac{\tilde{\mu}}{2} \hat{z} \times \nabla \eta , \quad V = \frac{1}{4} ( |\nabla \ln \tilde{\mu}|^2 + |\tilde{\mu} \nabla \eta|^2 ) - \frac{1}{2} \nabla^2 \ln \tilde{\mu} - \tilde{\mu} \varepsilon \omega^2.
\ee
Thus, the gyrotropic response produces an effective vector potential for light, which breaks time-reversal symmetry and can create topological band gaps with nonzero Chern number.

Unfortunately gytrotropic response is very weak at optical frequencies, severely limiting the size of any topological band gap. Therefore alternate approaches suitable for time reversal-symmetric dielectric materials are preferred. Synthetic gauge fields for light can also be introduced by periodic modulation of photonic crystal parameters in time or space. The effective magnetic field, associated with the complex phase accumulation, can thus force unidirectional propagation of photons without the use of magnetic materials at optical frequencies, in a way similar 
to electrons moving along skipping orbits along the edge in a two-dimensional magnetically-biased electron gas.  

For example, light propagation in waveguide lattices formed by a weak modulation $\delta n$ of the refractive index $n(\bs{r}) = \sqrt{\varepsilon(\bs{r})} = n_0 + \delta n(\bs{r})$ is described by the paraxial wave equation \citep{Rechtsman2013}
\be 
i \partial_z \psi(\bs{r}_{\perp},z) = -\frac{1}{2k_0} \nabla_{\perp}^2 \psi - \frac{k_0 \delta n(\bs{r}_{\perp},z)}{n_0} \psi (\bs{r}_{\perp},z),
\ee 
where $\bs{r}_{\perp} = (x,y)$ are the transverse coordinates, $z$ is the propagation direction, and $k_0 = 2\pi n_0 / \lambda$ is the wavelength of light in the medium. Since $z$ plays the role of ``time'' in the paraxial equation, broken inversion symmetry $\delta n(\bs{r}_{\perp},z) \neq \delta n(\bs{r}_{\perp}, -z)$ can be used to obtain nonzero Chern numbers for light propagating through the waveguide array.

The second most important class of topological phase is the quantum spin Hall or $Z_2$ topological insulator, which can be considered (loosely) as consisting of two copies of Chern topological insulators with gauge fields acting on opposite spins~\citep{Hasan2010}. 
In fact, it is the time-reversal symmetry that ensures the topological stability  
of the edge states supported by quantum spin Hall topological insulators, with spin-orbital coupling, in the absence of spin-flipping processes. However, one should keep in mind that while time-reversal symmetry alone is sufficient to guarantee the presence of degenerate spin-1/2 states in condensed-matter physics, owing to Kramers' theorem for fermions, this is not the case for photons because they obey the bosonic quantum statistics. Therefore, additional deliberately-engineered symmetries related to pseudo-time-reversal operators $\hat{\mathcal{C}}_{\text{b}}$, $\hat{\mathcal{C}}_{\text{b}}^2 = -1 $, $\hat{\mathcal{C}}_{\text{b}} \hat{\mathcal{H} } \hat{\mathcal{C}}_{\text{b}}^{-1} = \hat{\mathcal{H} } $ are required % for bosons 
in photonics to mimic the fermionic property of electrons $\hat{\mathcal{T}}_{\text{f}}^2=  -1$, with time-reversal operator $\hat{\mathcal{T}}_{\text{f}} = i \hat{\sigma}_y \hat{K}$, and achieve time-reversal-invariant topological order~\citep{Ozawa2019}. They include internal symmetry of the electromagnetic field inside a photonic structure or crystalline symmetries of a photonic lattice. Such symmetries can produce modal degeneracies in a band structure, generating doublet states to emulate the pseudospin degree of freedom, thus enabling photonic analogues of the quantum spin-Hall effect~\citep{Khanikaev2017}.

For example, metamaterials formed through subwavelength modifications to the material properties can be used to engineer constitutive relations not found in natural materials, including negative refractive indices and strong coupling between electric and magnetic field components. \cite{Khanikaev2013} exploited this tunability to engineer two-dimensional metamaterials with bianisotropic coupling between transverse electric (TE; $H_z$) and transverse magnetic (TM; $E_z$) field components,
\begin{align}
    \left( k_0^2 \mu_{zz} + \nabla_{\perp} \frac{1}{\varepsilon_{\perp}} \nabla_{\perp} \right) H_z &= \left[ \nabla_{\perp} \left( \frac{ -i \chi_{xy}} {\varepsilon_{\perp} \mu_{\perp}}\right) \times \nabla_{\perp} E_z \right]_z, \\
    \left( k_0^2 \varepsilon_{zz} + \nabla_{\perp} \frac{1}{\mu_{\perp}} \nabla_{\perp} \right) E_z &= \left[ \nabla_{\perp} \left( \frac{ -i \chi_{xy}} {\varepsilon_{\perp} \mu_{\perp}}\right) \times \nabla_{\perp} H_z \right]_z, 
\end{align}
By engineering the metamaterial parameters such that $\mu_{zz} = \varepsilon_{zz}$ and $\mu_{\perp} = \varepsilon_{\perp}$, the electric and magnetic field components become degenerate. Consequently, bianisotropic coupling $\chi_{xy} \neq 0$ then gives rise to new eigenmodes $\psi_{\pm} = E_z \pm H_z$ emulating the quantum spin Hall phase.

\section{Key issues}

\subsection{Are photonic quantum Hall effects genuinely robust?}
\label{sec:robustness}

The exact correspondence between electronic and photonic quantum Hall effects only holds under certain idealised conditions, including the absence of losses and optical nonlinearities. Remarkably, studies of the robustness of topological photonic edge states under non-ideal conditions have lead to the discovery of new kinds of topological phenomena in both photonic and condensed matter systems.

\subsubsection{Losses and non-Hermitian topological phases}

Early works on photonic topological edge states already recognized that unavoidable absorption, radiation, and scattering losses in photonic systems would place practical limits on their robustness and lifetime~\citep{Wang2009}. 

Material absorption is a fundamental property of optical media that cannot be eliminated via collective (band structure) effects. In fact, topological band structures typically emerge from confinement or multiple scattering, slowing down light propagation and enhancing absorption compared to homogeneous systems \citep{Baba2008}.

Radiation and scattering losses occur because there is no topological robustness of two-dimensional photonic quantum Hall effects against out-of-plane propagation. Modes of quasi-two-dimensional slabs with momenta above the light line will couple into the continuum of free space, resulting in radiative decay \citep{Gorlach2018}. Nominally-bound modes below the light line can still couple to radiation modes and acquire a finite lifetime in the presence of fabrication defect-induced scattering, such as roughness at boundaries between different materials comprising the slab \citep{PhysRevResearch.2.043109}. To avoid out-of-plane scattering it is necessary to embed the planar photonic crystal in a three-dimensional photonic crystal supporting a complete omnidirectional band-gap at the frequency of interest. 

Generally losses are manageable at microwave frequencies, where metals are good conductors and can provide perfect in-plane confinement and low-loss dielectric materials exist. At higher frequencies (terahertz and above) losses typically increase and feasible refractive index contrasts are weaker, making in-plane confinement harder. One way losses can be managed is by introducing optical gain (amplification), but this often brings about separate issues including amplification of noise and instabilities.

Regardless of their microscopic origin, the influence of loss and gain can be modelled by non-Hermitian effective Hamiltonians. Initial theoretical studies of the one-dimensional Su-Schrieffer-Heeger model suggested topologically-protected bulk \citep{PhysRevLett.102.065703} and edge \citep{Schomerus:13} phenomena could still persist in non-Hermitian systems, subsequently observed using optical waveguide arrays \citep{PhysRevLett.115.040402} and microwave resonator chains \citep{Poli2015}. Later studies have observed non-Hermitian topological phenomena in a variety of systems including photonic crystal slabs \citep{Zhen2015,doi:10.1126/science.aap9859} and microring resonator lattices \citep{doi:10.1126/science.aay1064}. \cite{RevModPhys.93.015005} provide a more comprehensive review of non-Hermitian topological band theory.

\subsubsection{Symmetry-breaking perturbations}

Even if losses can be perfectly eliminated or compensated, disorder and imperfections can still spoil the robustness of most photonic quantum Hall effects; only systems that break time-reversal symmetry can be genuinely robust to all kinds of (Hermitian) perturbations.

The time reversal-symmetric systems supporting spin Hall effects most widely used in photonics do not share the robustness of their electronic counterparts; the robustness of the electronic quantum spin Hall effect is due to backscattering suppression protected by the $\hat{\mathcal{T}}_{\text{f}}^2=  -1$ fermionic time-reversal symmetry. As noted in Sec.~\ref{sec:approaches}, the photonic time-reversal symmetry ordinarily obeys $\hat{\mathcal{T}}_{\text{b}}^2=  -1$; fermionic $\hat{\mathcal{T}}_{\text{f}}^2=  -1$ requires introduction of auxiliary spin-like degrees of freedom, such as valley or polarization \citep{Khanikaev2013}. Since this spin is not intrinsic but emerges due to structural properties, certain structural imperfections and lattice terminations can spoil the protection and induce backscattering \citep{PhysRevB.101.054307}.

For example, the valley Hall photonic crystals only support robust edge states in the absence of coupling between their two valleys. Short-scale structural deformations can induce inter-valley scattering that localizes the edge states. Similarly, topological crystalline phases such as the shrunken-expanded hexagonal photonic crystal design of \cite{Wu2015} are protected by crystalline rotational symmetries. Typically, the symmetry is broken locally at domain walls or edges, inducing small gaps in the spectra of the interface states \citep{Xu:16}.

Therefore topological robustness of symmetry-protected photonic quantum Hall effects only holds to the extent that the underlying symmetries can be preserved. This poses a challenge because fabrication disorders will typically not preserve any spatial symmetries.

\subsubsection{Optical nonlinearities}

What is remarkable about electronic quantum Hall effects is that, even though the basic physics can be captured by simple single particle band structures, they are robust to electron-electron interactions and other many-body effects, giving rise to robust quantized transport. It is thus interesting to ask whether similar robustness can hold for classical waves in the presence of nonlinear interactions. This is another frontier of studies of photonic quantum Hall effects, reviewed by \cite{Smirnova2020APR}.

Theoretical works have predicted that nonlinearities can lead to self-focusing \citep{PhysRevLett.117.143901}, instabilities \citep{PhysRevA.94.021801}, and nonlinear localization \citep{PhysRevLett.111.243905}, potentially destroying robust transport. Nonlinearities also induce coupling between different linear modes of the system, potentially channeling energy between robust edge states into non-robust bulk modes \citep{Xia2020}. On the other hand, nonlinear pumps can be used to break time-reversal symmetry \citep{PhysRevB.93.020502} and induce robust transport of weak probe beams \citep{PhysRevX.6.041026}.

Experimental studies on the effect of nonlinearities on photonic quantum Hall edge phases have observed a variety of effects, including frequency conversion \citep{Mittal2018,PhysRevLett.123.103901}, bulk soliton generation \citep{
doi:10.1126/science.aba8725}, nonlinearity-induced edge states \citep{doi:10.1126/science.abd2033, PhysRevX.11.041057}, and nonlinearity-induced higher order topological phases \citep{Kirsch2021}.

\begin{figure}
    \centering
    \includegraphics[width=\columnwidth]{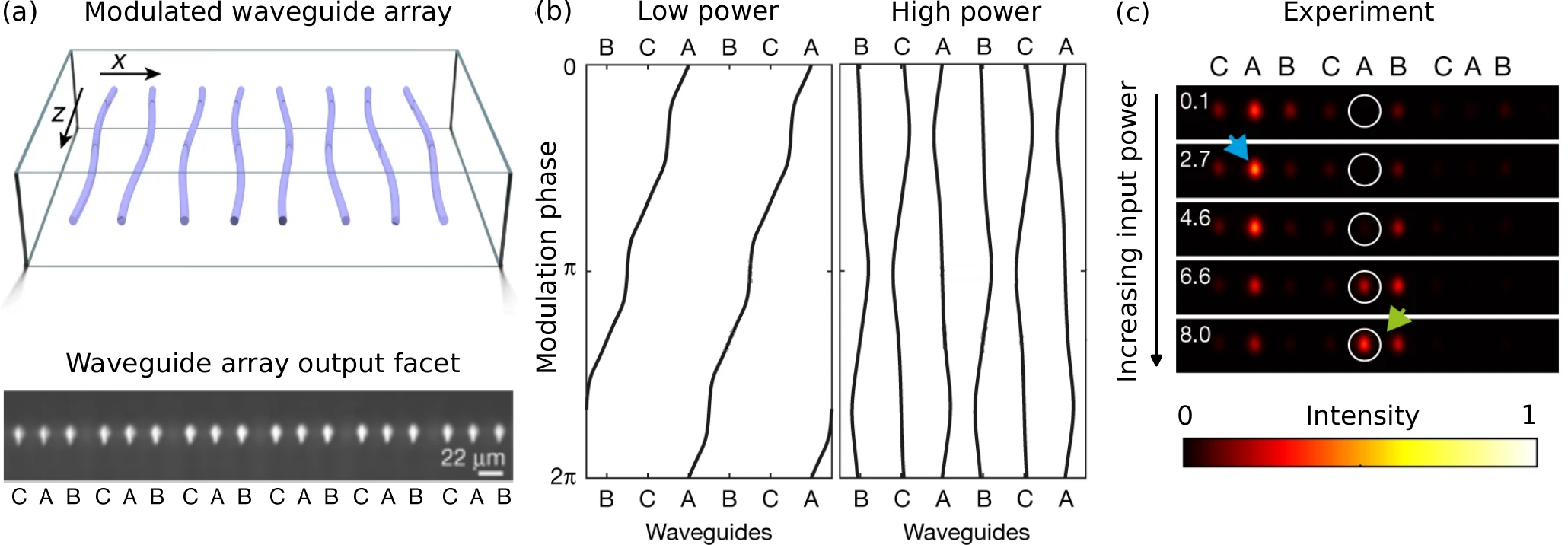}
    \caption{(a) One-dimensional periodically-modulated waveguide array implementing a Thouless pump. (b) Instantaneous nonlinear spectra during one modulation cycle. Low power solitons are pumped one unit cell per modulation cycle, whereas high power solitons remain pinned to the initial unit cell. (c) Experimental observation of quantized Thouless pumping of solitons. Low power output profile is displaced one unit cell with respect to the input waveguide (white circles), while output becomes pinned to input unit cell at high powers. Adapted from \cite{Jurgensen2021}.}
    \label{fig:Jurgensen2021}
\end{figure}

Since the dynamics of nonlinear systems cannot be completely characterized in terms of a unique spectrum or band structure, and are sensitive to the initial state and input intensity, rigorously demonstrating topological robustness in nonlinear wave systems is quite challenging. One notable example of where nonlinear robustness has been predicted and observed is in quantized Thouless pumping induced by periodic modulation of system parameters \citep{Jurgensen2021}, as shown in Fig.~\ref{fig:Jurgensen2021}. Following a complete modulation cycle the soliton solutions must be identical up to translation invariance, thus solitons must be pumped a quantized integer number of unit cells. This number can change as the input power is varied due to nonlinear bifurcations.

\subsection{Potential device applications}

Electronic quantum Hall effects are useful as the electrical resistance standard, thanks to their unexpectedly extreme robustness \citep{Klitzing2020}. On the other hand, photonic quantum Hall edge states have required clever and careful designs to observe, with claims of robustness typically based on the introduction of specific controlled perturbations. Will photonic quantum Hall effects also see useful applications? A few directions are currently being pursued.

\subsubsection{Topological waveguides and cavities}

Robust broadband non-reciprocal optical waveguiding was the first key application proposed for photonic quantum Hall effects \citep{Haldane2008}, however, practical implementations competitive with the best conventional waveguide designs remain out of reach, in part due to the limitations to topological robustness leading to enhanced losses and nonzero backscattering \citep{PhysRevResearch.2.043109} discussed in Sec.~\ref{sec:robustness}.

\begin{figure}
    \centering
    \includegraphics[width=0.9\columnwidth]{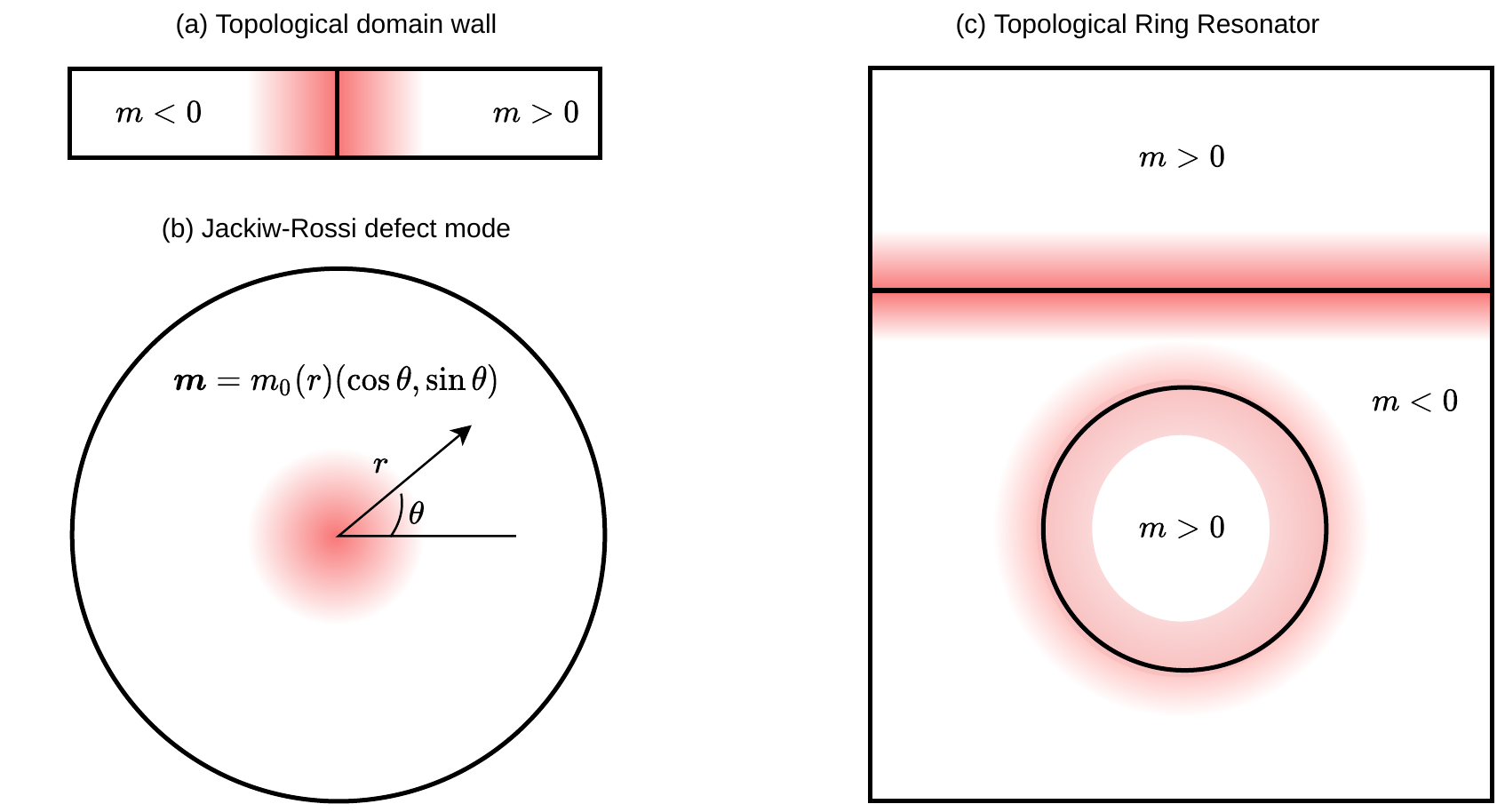}
    \caption{Examples of topological waveguides and cavities. (a) A domain wall between inequivalent topological phases (denoted by $m<0$ and $m>0$) results in a localized mid-gap mode in 1D systems an a unidirectional edge state in 2D systems. (b) Generation of robust 2D localized mode using a vortex-like modulation of the topological gap parameter $\boldsymbol{m}$. (c) Topological ring resonator coupled to a waveguide using topological domain walls in a 2D system.}
    \label{fig:cavities}
\end{figure}

Topological phases have also been considered as a way to create robust localized modes and resonators for light, with some examples illustrated in Fig.~\ref{fig:cavities}. The simplest approach is based on mid-gap localized modes appearing in one-dimensional photonic crystals \citep{Malkova2009}. Subsequently, topological resonators based on rings of valley Hall edge states \citep{https://doi.org/10.1002/lpor.201900087}, topological corner modes \citep{Mittal2019}, and topological vortex defect modes \citep{Gao2020} have been investigated. \cite{Kim2020} provide a more in-depth review.

One issue of vital importance to the design of waveguides and cavities is the device footprint. For example, in the case of two-dimensional quantum Hall edge states, robustness against backscattering is provided by the two-dimensional bulk. If one attempts to minimize the device footprint by reducing the size of the bulk, coupling between opposite edges will spoil the topological protection. Thus, topological waveguides and resonators tend to have a larger physical size (constrained by the size of the topological band gap) compared to conventional waveguides and resonators. This limitation needs to be taken into account when comparing the performance of topological and non-topological designs.

A more comprehensive overview on the design of topological photonic crystal waveguides and cavities is given by \cite{Iwamoto:21}.

\subsubsection{Topological lasers}

Lasers are arguably one of the most important inventions of the 20th century and are now ubiquitous in a range of communication, sensing, and fabrication technologies. Naturally there has been tremendous interest in studying whether photonic quantum Hall edge states or other topological phases can serve as the basis for new and improved laser designs with improved performance. 

\begin{figure}
    \centering
    \includegraphics[width=\columnwidth]{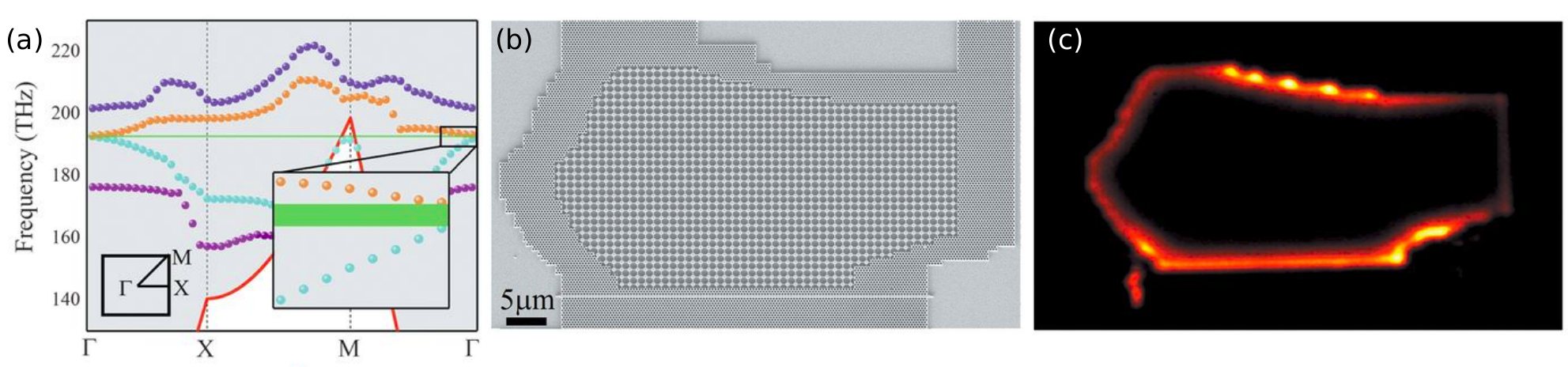}
    \caption{First two-dimensional topological laser. (a) Band structure of the InGaAsP-YIG photonic crystal. Under an applied magnetic field, the YIG gyrotropic response breaks time-reversal symmetry, opening a small topological band gap with nonzero Chern number (green shaded region in inset). (b) Topological cavity formed by irregular boundary between inner topological photonic crystal and outer trivial photonic crystal. (c) Image of the device under high power optical pumping, revealing emission from the topological edge mode. Adapted from \cite{doi:10.1126/science.aao4551}.}
    \label{fig:laser}
\end{figure}

The first experiments combining optical gain media with topological photonic systems considered mid-gap modes of one-dimensional lattices \citep{StJean2017,Zhao2018,Parto2018}. Shortly afterwards, \cite{doi:10.1126/science.aao4551} observed lasing of quantum Hall edge states using the magneto-optic photonic crystal shown in Fig.~\ref{fig:laser}, exploiting the fact that light emission from a continuous wave laser is intrinsically narrowband, which makes the weak magneto-optic response at optical frequencies sufficient to resolve localized edge states.

The first observation of lasing of quantum spin Hall states by \cite{Segev2018b} employed ring resonator lattices. Lasing in the spin Hall phase was observed to spontaneously break time-reversal symmetry, resulting in chiral light emission. Since lasing modes based on two-dimensional edge states appear to be insensitive to the shape of the boundary, they offer a potential route towards designing large cavities robust to fabrication disorders.

Since these seminal experiments numerous follow-up theoretical and experimental studies have been performed, considering different cavity designs and dimensions and various pumping schemes, reviewed by \cite{OtaTakataOzawaAmoJiaKanteNotomiArakawaIwamoto+2020+547+567}. Efforts now focus on two complementary directions: (1) How to maximize the modal volume while preserving stable single mode lasing, and (2) how to reduce the mode volume and increase the quality factor to design highly efficient, low power lasers.

The best route towards maximizing the modal volume remains under debate. Competing approaches include 2D quantum Hall-like edge states \citep{Segev2018b}, valley-Hall modes at domain walls \citep{Zeng2020}, bulk modes \citep{Shao2020}, vortex defect modes \citep{Gao2020,Yang2022}, and Dirac point modes of gapless band structures \citep{Contractor2022}. On the other hand, one obstacle against small modal volumes is that topological effects are a collective lattice property; protection is provided by having a sufficiently large bulk. It seems difficult to reduce the modal volumes below that achievable using a single isolated resonator or defect cavity; the main benefit of the topology may be in achieving a robustness to disorder at the expensive of a larger modal volume.

On the theory side, research on understanding the mechanisms for stable lasing in topological systems is ongoing. For example, it is commonly assumed that stable edge state lasing requires the optical gain to be localized to the edge of the system. On the other hand, even though the experiments shown in Fig.~\ref{fig:laser} employed uniform optical pumping, no lasing of bulk modes was observed. \cite{doi:10.1063/5.0041124} proposed a mechanism for this surprising observation based on excess gain occurring in the topologically-trivial cladding medium. Research on this topic including the interplay with non-Hermitian topological phenomena remains ongoing. 

\subsection{Making photonic quantum Hall effects quantum}

A long-standing challenge has been to integrate topological band structure effects with strong (single photon level) interactions to generate many-body bosonic quantum Hall effects. This is hard because nonlinear optical effects are typically very weak, requiring intricate schemes involving resonant nonlinearities \citep{PhysRevLett.79.1467}, ultra-high quality factor cavities, and pumping to compensate for losses. Comprehensive reviews on this subject and the quantum simulation of other strong interaction phenomena using photons are given by \cite{Noh2016} and \cite{Carusotto2020}.

\subsubsection{Theoretical proposals}

Early works proposed bosonic analogues of fractional quantum Hall states using atoms trapped in optical lattices and cavity arrays, using modulated couplings to induce effective magnetic fields for neutral particles \citep{PhysRevLett.94.086803,Hafezi2007,Cho2008,PhysRevLett.108.223602}. These proposals were then generalized to photonic systems with strong photon-photon interactions \citep{PhysRevA.82.043811,Nunnenkamp_2011,PhysRevA.84.043804,Hafezi2011}.

An important question specific to the photonic schemes is how to reliably generate (fractional) quantum Hall states. Early proposals predicted that in the presence of strong interactions and sufficiently weak losses, two-photon Laughlin states of light could be excited using a coherent pump tuned to the state's energy \citep{PhysRevLett.108.206809, Hafezi2013,PhysRevA.89.023803}.

However, uniform driving approaches cannot be easily scaled to larger systems where topological robustness is expected to emerge due to inevitable losses. Losses lead to spectral broadening of the modes, resulting in excitation of multiple modes in larger systems. Moreover, higher photon number states will be exponentially suppressed by losses. Subsequent studies proposed solutions based on two-photon driving designed to restore lost photons \citep{Kapit2014}, and local driving combined with topological pumping \cite{PhysRevLett.113.155301}.

\subsubsection{Experiments}

Experimentally, strong single photon-level interactions have been demonstrated at microwave frequencies using superconducting qubits \citep{PhysRevLett.113.220502} and at optical frequencies using Rydberg atoms \citep{Firstenberg2013}. In both these platforms, the main difficulty is to maintain the strong interactions while scaling up the system size and number of particles, while using fine-tuning to compensate for and minimize inevitable disorder effects.

\begin{figure}
    \centering
    \includegraphics[width=\columnwidth]{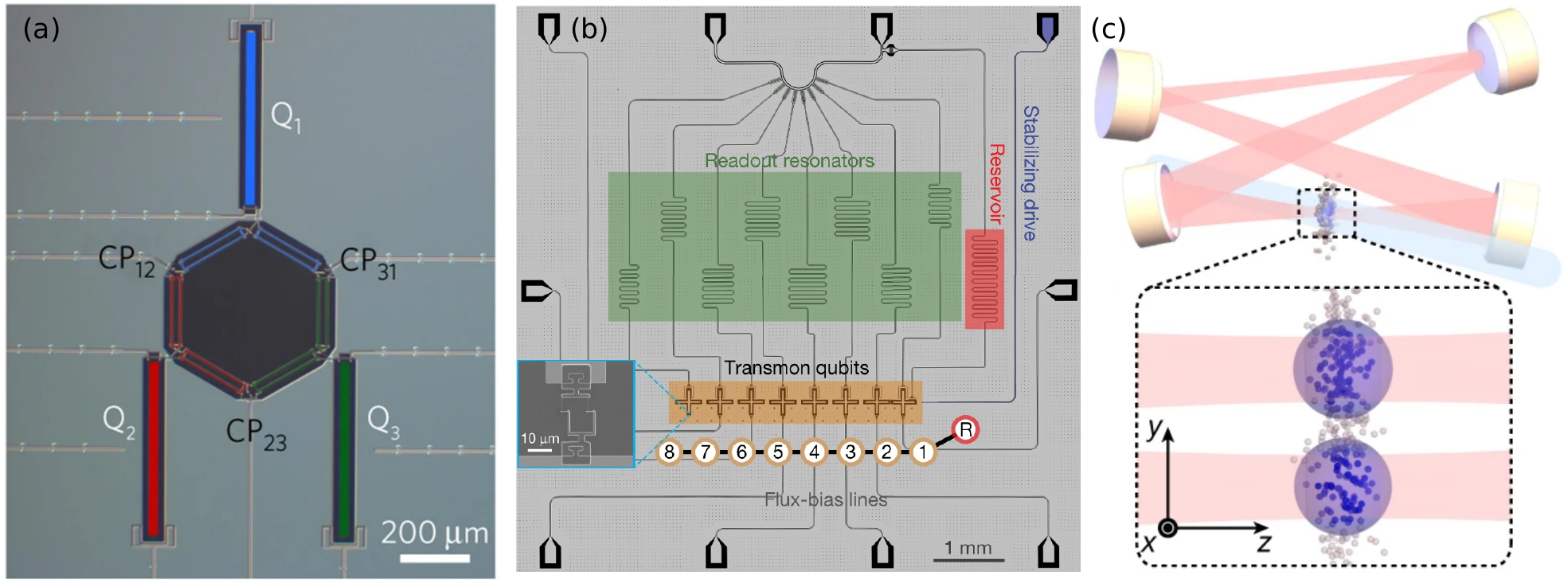}
    \caption{Experimental platforms for many-body bosonic quantum Hall effects. (a) Superconducting transmon qubits Q$_{i}$ connected via adjustable couplers CP$_{ij}$ which are periodically modulated in time to induce an effective magnetic flux, adapted from \cite{Roushan2017}. (b) Array of superconducting transmon qubits, where two-photon driving combined with coupling to a lossy auxiliary resonator reservoir is used to stabilize a photonic Mott insulator, adapted from \cite{Ma2019}. (c) Twisted four-mirror optical cavity used to generate photonic Landau levels, with strong photon-photon interactions mediated using trapped Rydberg atoms (inset), adapted from \cite{Clark2020}.}
    \label{fig:quantum}
\end{figure}

\cite{Roushan2017} were the first to combine strong interactions with a synthetic magnetic field using the three-site lattice of superconducting microwave qubits shown in Fig.~\ref{fig:quantum}(a), observing a photon number-dependent chirality of the dynamics. Subsequent work observed a Hofstadter butterfly spectrum in a quasiperiodically-modulated one-dimensional lattice \citep{doi:10.1126/science.aao1401}. \cite{Ma2019} implemented two-photon driving to prepare a Mott insulator state of photons in the eight-site lattice shown in Fig.~\ref{fig:quantum}(b). Implementation of fractional quantum Hall states in this platform remains an outstanding challenge, requiring the combination of a two-dimensional lattice geometry with precise control over the qubit couplings and frequencies.

Parallel efforts targeted at optical frequencies started by generating a continuum Landau level spectrum for photons using the twisted ring resonator \citep{Schine2016} shown in Fig.~\ref{fig:quantum}(c). This platform can be combined with strong single photon-level interactions by placing ultra-cold Rydberg atoms within the resonator \citep{Clark2019}, enabling generation of two-photon Laughlin states of light \citep{Clark2020}, detected by measuring angular two-photon correlations. Future work aims to combine this platform with dissipative preparation schemes to assemble many-body quantum Hall states of light.

\section{Summary and Future Directions}

Over the past decade the emulation of topological phenomena in condensed matter physics including quantum Hall effects has captured the imagination of researchers in photonics. In this Chapter we have attempted to provide a broad overview of this active and still-developing field from its early origins up to topics of ongoing research. While it remains unclear as to whether photonic quantum Hall effects will remain a curiosity among researchers or will form the basis for new and improved devices, there is no doubt that this elegant formalism completes the band theory describing periodic photonic media incluing photonic crystals \citep{JoannopoulosBook}.

There are numerous future directions for research. On the fundamental side, there are ongoing efforts towards understanding the properties of recently-discovered topological phenomena including higher order topological phases, topological defect modes, non-Hermitian topological phases, and the role of topology in nonlinear and strongly-interacting systems, as well as observing these phenomena in experiments. At the same time, one (valid) criticism of the ongoing tremendous interest in topological photonic systems is that they are exotica; a solution in search of a problem. Existing approaches towards generating photonic quantum Hall effects have been largely based on clever and intricate designs. To realize the dream of useful device applications we need to develop new approaches to achieve improved figures of merit compared to conventional design approaches.

Photonic quantum Hall effects are one of many examples of how ideas from other research fields can inspire new approaches towards the design and understanding of photonic systems. It is natural to ask where the next big inspiration will come from. One leading candidate is machine learning, a topic which has seen tremendous progress over the past decade and has now started to capture the attention of physicists \citep{RevModPhys.91.045002}. Not only can ideas from machine learning be applied towards analysis of photonic systems, but concepts from topological photonics might be useful as a means for designing photonic hardware for machine learning including neuromorphic computing \citep{Tan2021} and image processing \citep{PhysRevApplied.11.054033, PhysRevApplied.11.034043, Zhu_NatCom21} with ultrafast speed and low energy consumption.  

% \bibliographystyle{agsm1}
% \bibliography{TheNewBibliography}

\end{document}